\documentclass[aps,preprints,preprintnumbers,showpacs,superscriptaddress,groupedaddress,nofootinbib]{revtex4}  
\usepackage{graphicx}  
\usepackage{dcolumn}   
\usepackage{bm}        
\usepackage{amssymb}   
\usepackage[usenames, dvipsnames]{color}

\usepackage{psfrag}
\usepackage{epsfig}
\usepackage{epstopdf}
\usepackage{slashed}
\usepackage[colorlinks, hyperindex]{hyperref}
	\hypersetup
	{
		colorlinks, %
		citecolor=blue, %
		linkcolor=red, %
		urlcolor=blue, %
	}

\usepackage{CJK}

\begin{document}
\begin{CJK*}{UTF8}{gbsn}

\title{Leading order relativistic chiral nucleon-nucleon interaction}

\author{Xiu-Lei~Ren}
\affiliation{State Key Laboratory of Nuclear Physics and Technology, School of Physics, Peking University, Beijing 100871, China}
\affiliation{Institut f\"{u}r Theoretische Physik II, Ruhr-Universit\"{a}t Bochum, D-44780 Bochum, Germany}

\author{Kai-Wen~Li}
\affiliation{School of Physics and
Nuclear Energy Engineering \& International Research Center for Nuclei and Particles in the Cosmos, Beihang University, Beijing 100191, China}

\author{Li-Sheng~Geng}
\email[]{Email: lisheng.geng@buaa.edu.cn}
\affiliation{School of Physics and
Nuclear Energy Engineering \& International Research Center for Nuclei and Particles in the Cosmos, Beihang University, Beijing 100191, China}
\affiliation{Beijing Key Laboratory of Advanced Nuclear Materials and Physics, Beihang University, Beijing 100191, China}

\author{Bingwei Long}
\affiliation{Center for Theoretical Physics, Department of Physics, Sichuan University, Chengdu, Sichuan 610064, China}

\author{Peter~Ring}
\affiliation{Physik Department, Technische Universit\"{a}t M\"{u}nchen, D-85748 Garching, Germany}
\affiliation{State Key Laboratory of Nuclear Physics and Technology, School of Physics, Peking University, Beijing 100871, China}

\author{Jie~Meng}
\email[]{Email: mengj@pku.edu.cn}
\affiliation{State Key Laboratory of Nuclear Physics and Technology, School of Physics, Peking University, Beijing 100871, China}
\affiliation{School of Physics and
Nuclear Energy Engineering \& International Research Center for Nuclei and Particles in the Cosmos, Beihang University, Beijing 100191, China}

\date{\today}

\begin{abstract}
Motivated by the successes of relativistic theories in studies of atomic/molecular and nuclear systems and the  need for
a relativistic chiral force in relativistic nuclear structure studies, we explore a new relativistic scheme to
construct the nucleon-nucleon interaction in the framework of covariant chiral effective field theory. The chiral
interaction is formulated up to leading order with covariant power counting and a Lorentz invariant chiral Lagrangian.
We find that the relativistic scheme induces all six spin operators needed to describe the nuclear force. A
  detailed investigation of the partial wave potentials shows a better description of the $^1S_0$ and $^3P_0$ phase
  shifts than the leading order Weinberg approach, and similar to that of the next-to-leading order Weinberg
  approach. For the other partial waves with angular momenta $J\geq 1$, the relativistic results are almost the same as their leading order non-relativistic counterparts.
\end{abstract}

\maketitle

\section{Introduction}

 There is strong evidence that relativistic effects play an
indispensable role in our understanding of the fine structure of atoms/molecules~\cite{Schwerdtfegerbook} and nuclei~\cite{Mengbook}, although non-relativistic methods were historically very popular and
are still routinely utilized in modern studies. The most familiar manifestations of relativistic effects include the appearance of anti-fermions, their spin and the resulting spin-orbit interactions, which form a key to understand the spin-orbit splitting of atomic spectra and nuclear single particle levels~\cite{mayer1955elementary}. In contrast to kinematical effects, which at low energies can often be neglected or treated perturbatively, these are dynamical effects, in particular the velocity dependent potentials such as the
spin-orbit force. Today, studies of complex atomic/molecular systems have reached a high level of maturity~\cite{Nobel2013Chem}, while similar studies of nuclear structure and reactions are still at an early stage~\cite{Elhatisari:2015iga}. 

There are two important differences between these two systems.
A key difference in microscopic studies of atoms/molecules and nuclei, though they share similar theoretical approaches, is the dominating fundamental interaction. For atoms/molecules, the electromagnetic force is known rather accurately both at the classical level and at the field theoretical level. On the other hand, for nuclei, the nuclear force, being a residual interaction of the strong force, is still far from being completely understood (see, e.g., Ref.~\cite{Lapoux:2016exf}). There is, however, a second important difference. In atoms and molecules the electromagnetic force is a
Lorentz-vector and as a consequence the Coulomb potential also causes spin-orbit 
splitting. The nuclear force, however, contains extremely large Lorentz 
scalars and Lorentz vectors of opposite sign. This is a direct consequence of
QCD, as has been shown by Cohen {\it et al.}~\cite{Cohen:1991js,Cohen1992_PRC45-1881,Cohen1995_PPNP35-221}. Scalar and vector forces
cancel to a large extent in the normal potential, but they add up in the spin-orbit term. 
In Coulombic systems, the velocity dependent spin-orbit term is small and
in many cases (even in high precision calculations) it is treated perturbatively. This is possible, because the velocities are not large.
In nuclei the velocities are not large either, but, in the velocity dependent terms, 
the factor in front of the velocity is very large and forbids a perturbative treatment.

After more than 80 years of extensive study since the pion-exchange picture was proposed~\cite{Yukawa:1935xg}, the nuclear force still remains a central topic in nuclear physics and nuclear astrophysics. There are
a variety of formulations of the nuclear force.
Most studies are performed in the non-relativistic (NR) framework, including the high precision phenomenological nuclear potentials, Reid93~\cite{Stoks:1994wp} and Argonne $V_{18}$~\cite{Wiringa:1994wb},  or the chiral forces~\cite{Epelbaum:2014sza,Entem:2017gor}. In the relativistic framework, only two formulations have been studied rather extensively, namely the (CD-)Bonn potential~\cite{Machleidt:1989tm,Machleidt:2000ge} and the covariant spectator theory~\cite{Gross:1991pm,Gross:2008ps}~\footnote{We note that a covariant calculation of two-pion exchanges exists \cite{Higa:2003jk, Higa:2007gz} using the infrared regularization~\cite{Becher:1999he}.}. Of these, the Bonn potential has been successfully applied in relativistic many-body calculations (e.g. Dirac-Brueckner-Hartree-Fock theory~\cite{TerHaar:1986xpv}), to study both nuclear matter~\cite{Brockmann:1990cn} and, more recently, finite nuclei~\cite{Shen:2016bva,Shen:2017vqr}. However, the connection of the relativistic phenomenological potentials to the underlying theory of the strong interaction, Quantum Chromodynamics (QCD), is not very transparent. In this regard, a relativistic nucleon-nucleon interaction based on chiral effective field theory (ChEFT) is indispensable.

As a low energy effective field theory of non-perturbative QCD~\cite{Weinberg:1978kz}, ChEFT provides a model independent approach to study strong-interaction phenomena. It has been successfully applied to the mesonic sector and to systems involving baryons and heavy (flavored) hadrons.
Due to the large non-zero mass of the nucleon (compared to the pion mass), in the latter systems,
conventionally NR ChEFT, i.e. the heavy baryon (HB) scheme~\cite{Jenkins:1990jv}, is often used, especially for the two-baryon (few-body) sectors. In the 1990s,
Weinberg proposed to construct two-(few)-body interactions
  from chiral Lagrangians. First, one calculates the irreducible diagrams  in HB ChEFT perturbatively, and then
  uses the Lippmann-Schwinger equation to obtain transition amplitudes~\cite{Weinberg:1990rz,Weinberg:1991um}. This
  method is conventionally referred to as the Weinberg approach.
  Since then, the nucleon-nucleon interaction has been extensively
  investigated~\cite{Ordonez:1992xp,Ordonez:1993tn,vanKolck:1994yi,Kaiser:1997mw,Kaiser:1999ff,Kaiser:2001at,Kaiser:2001pc,
    Epelbaum:1998ka,Epelbaum:1999dj,Entem:2001cg,Entem:2002sf,Entem:2003ft,Epelbaum:2004fk,Epelbaum:2014efa}
  (see the reviews in Refs.~\cite{Bedaque:2002mn,Epelbaum:2008ga,Machleidt:2011zz} and references therein). Recently,
  chiral {\it NN} forces have  been constructed up to the fifth order by the Bochum-Juelich~\cite{Epelbaum:2014sza} and
    the Idaho groups~\cite{Entem:2014msa,Entem:2017gor}. The dominant two- and three-pion-exchange contributions at the sixth order have also been worked out, in Ref.~\cite{Entem:2015xwa}.
However, the Weinberg power counting scheme has been found to be non-renormalizable~\cite{Savage:1998vh}. To cure this,  several possible approaches have
been proposed~\cite{Kaplan:1998tg,Kaplan:1998we,Gegelia:1998ee,Cohen:1998jr,Gegelia:1999ja,Fleming:1999ee,Frederico:1999ps,Beane:2000wh,Beane:2001bc, Nogga:2005hy,Birse:2005um,Timoteo:2005ia,Birse:2007sx,Yang:2007hb,Yang:2009kx,Yang:2009pn,
Valderrama:2009ei,Valderrama:2011mv,Long:2011qx,Long:2012ve,Epelbaum:2012ua,Epelbaum:2015sha}, but the problem has not yet been fully resolved.

Meanwhile, in recent years, covariant ChEFT has been shown to be able to solve a number of
long-standing issues. It has shown relatively faster convergence than its NR counterpart in the one-baryon
sector~\cite{Geng:2008mf,Ren:2012aj,Ren:2014vea,Blin:2015era,Yao:2016vbz} and in heavy-light systems~\cite{Altenbuchinger:2013vwa}.
In addition to being covariant it satisfies analyticity constraints (for a short
review see Ref.~\cite{Geng:2013xn}). Motivated by these successes and the demand in relativistic nuclear structure studies, we explore a covariant
power counting scheme, which keeps the small components of Dirac spinors, to construct, in the framework of ChEFT, a relativistic {\it NN} potential in analogy to the phenomenological Bonn potential. As a first step, we investigate in this paper the possibility of constructing such a chiral force  up to leading order.
In the long run, however, we aim to also include  higher orders and to provide a high-precision relativistic chiral nuclear force so that relativistic
many-body calculations, such as those of Refs.~\cite{Brockmann:1990cn,Shen:2016bva,Shen:2017vqr} using
Dirac-Brueckner-Hartree-Fock theory, can be performed with these relativistic chiral forces.

The covariant power counting scheme presented here and also the main purpose of this investigation are quite different from those of Ref.~\cite{Epelbaum:2012ua}, where relativistic effects are, for the first time, included in a perturbative way, to derive a chiral force applicable to NR  calculations with particular focus on renormalization group invariance.

In this work, we start from a manifestly Lorentz invariant chiral Lagrangian and construct a relativistic
chiral nuclear force up to leading order (LO). To account for the non-perturbative nature of the
nucleon-nucleon interaction, we use a relativistic three-dimensional reduction of the Bethe-Salpeter equation,
as conventionally done by the nuclear structure community, to obtain the scattering amplitude from the chiral potential.
By fitting to the Nijmegen partial wave phase shifts, it is shown that one can achieve a satisfactory description of the
phase shifts of low angular momenta even at LO.

\section{Theoretical framework}

\subsection{Definition of potentials}
The concept of potentials is often used in the non-relativistic Schr\"{o}dinger and Lippmann-Schwinger equations. Since we are now working in covariant chiral EFT, it is worth clarifying the definition of potentials from a field-theoretical point of view.
Such a concept has already been thoroughly discussed in the 1970s (see, e.g.,  Refs.~\cite{Partovi:1969wd,Erkelenz:1974uj}), namely that the interaction field Hamiltonian appearing in a relativistic three-dimensional dynamical equation can be referred to as a two-nucleon potential.
To keep the manuscript self-contained, we would like to show the main procedures to introduce the potential in our relativistic framework.
For nucleon-nucleon elastic scattering, the ladder Bethe-Salpeter equation in operator form reads as
\begin{equation}
  \mathcal{T}(p',p|W) = \mathcal{A}(p',p|W) + \int \frac{d^4 k}{(2\pi^4)} \mathcal{A}(p',k|W) G(k|W) \mathcal{T}(k,p|W),
\end{equation}
where $p$ ($p'$) is the initial (final) relative four-momentum in the center-of-momentum system, and $W=(\sqrt{s}/2,\bm{0})$
is half of the total four-momentum with the total energy $\sqrt{s}=2E_p=2E_{p'}$ and $E_p=\sqrt{\bm{p}^2+m_N^2}$.
$\mathcal{T}$ denotes the invariant amplitude, and $\mathcal{A}$ is the interaction kernel consisting of all irreducible
diagrams appearing in covariant ChEFT. The free two-nucleon Green function reads
\begin{equation}
  G(k|W) = \frac{i}{[\gamma^\mu (W+k)_\mu - m_N + i\epsilon]^{(1)}~[\gamma^\mu (W-k)_\mu - m_N + i\epsilon]^{(2)}},
\end{equation}
where the superscripts refer to particles (1) and (2). The spin and isospin indices are suppressed.
However, in the low-energy region of the two-nucleon system, this is difficult to implement in practice, and a three-dimensional (3D) reduced equation is often used.
The reduction procedure is to replace $G$ by a three-dimensional $g$ which can produce the analytic structure of $G$
in the physical region only. In principle, there are infinite choices of $g$~\cite{Yaes:1971vw}. Nowadays, the
commonly used 3D reduced equations are, e.g., the Thompson equation~\cite{Thompson:1970wt}, the Blankenbecler-Sugar
equation~\cite{Blankenbecler:1965gx}, the Kadyshevsky equation~\cite{Kadyshevsky:1967rs}, or the Gross equation~\cite{Gross:1969rv}.  (Please refer to Ref.~\cite{Woloshyn:1974wm} for a comparison of different 3D relativistic scattering equations). In this work, we employ the Kadyshevsky equation, as shown in Refs.~\cite{Epelbaum:2012ua,Li:2016paq}. The corresponding Green function $g$ is
\begin{equation}
  g= 2\pi \frac{m_N^2}{E_k^2}\frac{\Lambda_+^{(1)}(\bm{k}) \Lambda_+^{(2)}(-\bm{k})}{\sqrt{s}-2E_k+i\epsilon} \delta[k_0-(E_k-\frac{1}{2}\sqrt{s})],
\end{equation}
where $E_k=\sqrt{\bm{k}^2+m_N^2}$ and $\Lambda_+^{(i)}$ ($i=1,~2$) are the positive energy projection operators for the two intermediate nucleons.
Using $G=g+(G-g)$, one can  rewrite  the Bethe-Salpeter equation schematically as two coupled equations,
\begin{eqnarray}
  \mathcal{T} &=& \mathcal{V} + \mathcal{V} g \mathcal{T},\label{Eq:3Dsym}\\
  \mathcal{V} &=& \mathcal{A} + \mathcal{A} (G-g) \mathcal{V}, \label{Eq:mathcalV}
\end{eqnarray}
where $\mathcal{V}$ is an the effective interaction kernel. After integrating out the time component $k_0$, Eq.~(\ref{Eq:3Dsym}) becomes a three-dimensional integral equation,
\begin{eqnarray}\label{Eq:4to3}
  && \mathcal{T}[p'_0=E_{p'}-1/2\sqrt{s},\bm{p}';p_0=E_{p}-1/2\sqrt{s},\bm{p}|W] = \mathcal{V}[p'_0=E_{p'}-1/2\sqrt{s},\bm{p}';p_0=E_{p}-1/2\sqrt{s},\bm{p}|W] \nonumber\\
  &&\quad + \int \frac{d^3k}{(2\pi)^3}~ \mathcal{V}[p'_0=E_{p'}-1/2\sqrt{s},\bm{p}';k_0=E_{k}-1/2\sqrt{s},\bm{k}|W] \nonumber\\
  &&\qquad\times \frac{m_N^2}{E_k^2}\frac{\Lambda_+^{(1)}(\bm{k}) \Lambda_+^{(2)}(-\bm{k})}{\sqrt{s}-E_k+i\epsilon}~\mathcal{T}[k_0=E_{k}-1/2\sqrt{s},\bm{k}; p_0=E_{p}-1/2\sqrt{s},\bm{p}|W].
\end{eqnarray}
We restrict the elements of $\mathcal{T}$ connecting the positive-energy spinors.
After sandwiching Eq.~(\ref{Eq:4to3}) between the Dirac spinors~$u(\bm{p},s)$, one obtains the $T$ matrix elements for {\it NN} scattering
\begin{equation}\label{eq:kadyshevsky}
   T(\bm{p}', \bm{p}) = V(\bm{p}', \bm{p}) + \int \frac{d^3 k}{(2\pi)^3} ~V(\bm{p}', \bm{k}) ~ \frac{m_N^2}{2E_k^2} \frac{1}{E_p-E_k + i\epsilon} T(\bm{k}, \bm{p}),
\end{equation}
where $V$ is our potential, defined as
\begin{equation}\label{Eq:DefineV}
  V(\bm{p}', \bm{p}) = \bar{u}(\bm{p}',s_1) \bar{u}(-\bm{p}',s_2) \mathcal{V}[p'_0=E_{p'}-1/2\sqrt{s},\bm{p}';p_0=E_{p}-1/2\sqrt{s},\bm{p}|W] u(\bm{p}, s_1) u(\bm{p}', s_2).
\end{equation}
The effective interaction kernel $\mathcal{V}$, determined by Eq.~(\ref{Eq:mathcalV}), can be perturbatively calculated via
\begin{eqnarray}\label{Eq:mathcalVpert}
  \mathcal{V}^{(0)} &=& \mathcal{A}^{(0)},\nonumber\\
  \mathcal{V}^{(2)} &=& \mathcal{A}^{(2)} + \mathcal{A}^{(0)} (G-g) \mathcal{A}^{(0)},
\end{eqnarray}
and so on. The superscripts refer to the order of chiral dimension of a particular Feynman diagram defined in Eq.~(11). Therefore, in a covariant formulation of ChEFT, one can obtain the potential $V(\bm{p}', \bm{p})$ defined in Eq.~(\ref{Eq:DefineV}) with the help of Eq.~(\ref{Eq:mathcalVpert}) , where the interaction kernel $\mathcal{A}$ is the sum of all the irreducible diagrams at a certain order.

%
%

\subsection{Leading order potential from covariant ChEFT}

In the relativistic framework, we retain the full form of Dirac spinors,  which have the usual form
\begin{equation}
  u(\bm{p},s) = N_p
                 \left(\begin{array}{c}
                        1 \\
                        \frac{\bm{\sigma}\cdot\bm{p}}{\epsilon_p}
                 \end{array}\right)\chi_{s}, \quad  N_p=\sqrt{\frac{\epsilon_p}{2M_N}},
\end{equation}
with $\epsilon_p = E_p+M_N$ and the Pauli spinor $\chi_s$.
A covariant power counting is tentatively introduced, which  uses naive dimensional analysis to determine the chiral dimension $(n_\chi)$ of a Feynman diagram with $L$ loops as
\begin{equation}
  n_\chi = 4 L - 2N_\pi - N_n + \sum\limits_k k V_k,
\end{equation}
where $N_{\pi}$ $(N_{n})$ is the number of internal pion (nucleon) propagators, and $V_k$ is the number of vertices from $k$th-order Lagrangians. The small expansion parameter in the covariant power counting is
the pion mass or the three-momentum of the nucleon. We would like to point out that the current covariant power counting is well defined in the $\pi\pi$ and $\pi N$ sectors, while for the $NN$ sector, such a power counting is not yet systematically formulated up to higher orders. Currently, we follow the arguments of Refs.~\cite{Djukanovic:2007zz,Girlanda:2010ya}, where the chiral dimension of effective Lagrangians for contact terms, beyond leading order, is determined by the partial derivatives on nucleon fields.

\begin{figure}[b]
  \centering
  \includegraphics[width=0.25\textwidth,keepaspectratio,angle=0,clip]{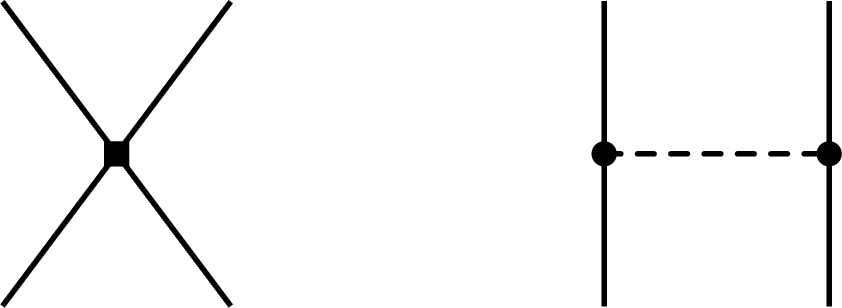}
  \caption{Feynman diagrams contributing to the nucleon-nucleon interaction at leading order in the covariant power counting. Solid lines denote nucleons and the dashed line represents the pion. The box denotes the vertex from $\mathcal{L}_{NN}^{(0)}$, while the dots show vertices from $\mathcal{L}_{\pi N}^{(1)}$.
  \label{Fig:FeynLO}
  }
\end{figure}

According to the above power counting, at leading order one needs to compute the Feynman diagrams shown in  Fig.~\ref{Fig:FeynLO}.
The relevant   Lagrangians are
\begin{equation}
  \mathcal{L}_\mathrm{eff.} = \mathcal{L}_{\pi\pi}^{(2)} + \mathcal{L}_{\pi N}^{(1)} + \mathcal{L}_{NN}^{(0)},
\end{equation}
where the superscript denotes  the chiral dimension. The lowest order $\pi\pi$ and $\pi N$ Lagrangians read,
\begin{eqnarray}
  \mathcal{L}_{\pi\pi}^{(2)} &=& \frac{f_\pi^2}{4} \mathrm{Tr}\left[\partial_\mu U \partial^\mu U^\dag + (U+U^\dag) m_\pi^2\right],\\
  \mathcal{L}_{\pi N}^{(1)} &=& \bar{\Psi}\left[i\slashed D - M_{N} + \frac{g_A}{2}\gamma^\mu \gamma_5 u_\mu \right]\Psi,
\end{eqnarray}
with the pion decay constant $f_\pi=92.4$ MeV, the axial vector coupling $g_A=1.267$~\cite{Olive:2016xmw}, and the SU(2)
matrix $U=u^2=\mathrm{exp}\left(\frac{i\Phi}{f_\pi}\right)$, where $\Phi$ and $\Psi$ contain the pion and nucleon fields,
\begin{equation}
  \Phi = \left(\begin{array}{cc}
           \pi^0 & \sqrt{2}\pi^+ \\
             \sqrt{2}\pi^- & -\pi^0
          \end{array}
  \right), \quad    \Psi = \left(
                 \begin{array}{c}
                   p \\
                   n \\
                 \end{array}
               \right).
\end{equation}
The covariant derivative of $\Psi$ is defined as
\begin{eqnarray}
  D_\mu \Psi &=& \partial_\mu \Psi + [\Gamma_\mu, \Psi],\\
  \Gamma_\mu &=& \frac{1}{2}\left(u^\dag\partial_\mu u + u \partial_\mu u^\dag\right),
\end{eqnarray}
and the axial current $u_\mu$ is
\begin{equation}
  u_\mu = i\left(u^\dag \partial_\mu u - u \partial_\mu u^\dag \right).
\end{equation}

The covariant four-fermion contact terms are provided by the following Lagrangian~\cite{Polinder:2006zh,Girlanda:2010ya,Djukanovic:2007zz},
\begin{eqnarray}
  \mathcal{L}_{NN}^{(0)} &=& \frac{1}{2} \left[C_S (\bar{\Psi}\Psi) (\bar{\Psi}\Psi) + C_A (\bar{\Psi}\gamma_5\Psi) (\bar{\Psi}\gamma_5\Psi)\right.  \nonumber\\
  &+& C_V (\bar{\Psi} \gamma_\mu \Psi) (\bar{\Psi} \gamma^\mu\Psi) + C_{AV} (\bar{\Psi} \gamma_\mu \gamma_5\Psi) (\bar{\Psi}\gamma^\mu\gamma_5\Psi) \nonumber\\
  &+& \left. C_T (\bar{\Psi}\sigma_{\mu\nu}\Psi) (\bar{\Psi}\sigma^{\mu\nu}\Psi)\right],
\end{eqnarray}
where $C_{S,A,V,AV,T}$ are low-energy constants (LECs). The $C_A$ term is considered of higher order by
some authors because it connects large and small components of the Dirac spinors~\cite{Petschauer:2013uua}. In our
case, we do not expand the Dirac spinors and therefore retain it. Explicit numerical studies show that this term plays a relatively minor role, however.

Since at the lowest order, $V_\mathrm{LO}=\bar{u}_1\bar{u}_2 \mathcal{A}_\mathrm{LO} u_1 u_2$, one can easily obtain the relativistic potential $V$, which is the sum of a contact term and a one-pion-exchange diagram,
\begin{equation}\label{Eq:covLO}
  V_\mathrm{LO} = V_\mathrm{CTP} + V_\mathrm{OPEP},
\end{equation}
where the contact potential (CTP) is
\begin{eqnarray}
  V_\mathrm{CTP}(\bm{p}',\bm{p}) &=& C_S \left(\bar{u}(\bm{p}', s_1') u(\bm{p}, s_1)\right) \left(\bar{u}(-\bm{p}', s_2') u(-\bm{p}, s_2)\right) \nonumber\\
    &+& C_A \left(\bar{u}(\bm{p}', s_1') \gamma_5 u(\bm{p}, s_1)\right) \left(\bar{u}(-\bm{p}', s_2') \gamma_5 u(-\bm{p}, s_2)\right) \nonumber\\
    &+& C_V \left(\bar{u}(\bm{p}', s_1') \gamma_\mu u(\bm{p}, s_1)\right) \left(\bar{u}(-\bm{p}', s_2') \gamma^\mu u(-\bm{p}, s_2)\right) \nonumber\\
    &+& C_{AV}\left(\bar{u}(\bm{p}', s_1') \gamma_\mu\gamma_5 u(\bm{p}, s_1)\right) \left(\bar{u}(-\bm{p}', s_2') \gamma^\mu\gamma_5 u(-\bm{p}, s_2)\right)  \nonumber\\
    &+& C_{T} \left(\bar{u}(\bm{p}', s_1') \sigma_{\mu\nu} u(\bm{p}, s_1)\right) \left(\bar{u}(-\bm{p}', s_2') \sigma^{\mu\nu} u(-\bm{p}, s_2)\right) ,
\end{eqnarray}
and the one-pion-exchange potential (OPEP) is,
\begin{equation}
  V_\mathrm{OPEP}(\bm{p}',\bm{p}) = -\frac{g_A^2}{4f_\pi^2} \frac{\left(\bar{u}(\bm{p}', s_1')\bm{\tau}_1\gamma^\mu\gamma_5q_\mu u(\bm{p}, s_1)\right)\cdot
 \left(\bar{u}(-\bm{p}', s_2') \bm{\tau}_2\gamma^\nu\gamma_5q_\nu u(-\bm{p}, s_2)\right)}
 {(E_{p'}-E_p)^2-(\bm{p}'-\bm{p})^2-m_\pi^2},
\end{equation}
where $q$ represents the four momentum transferred, $q=(E_{p'}-E_p,\bm{p}'-\bm{p})$, and $\bm{\tau}$ are the isospin Pauli matrices.

%

Expressing $V_\mathrm{LO}$ in terms of the Pauli matrices, one can easily see that the relativistic contact and OPE potentials contain all six  spin operators needed to describe the nuclear force~\cite{Goto1961},
\begin{eqnarray}
&&1,~~ \bm{\sigma}_1\cdot\bm{\sigma}_2,~~ \frac{i}{2}(\bm{\sigma}_1+\bm{\sigma}_2)\cdot(\bm{k}\times\bm{q}), ~~\bm{\sigma}_1\cdot\bm{q}\bm{\sigma}_2\cdot\bm{q},\nonumber\\
&& \bm{\sigma}_1\cdot\bm{k}\bm{\sigma}_2\cdot\bm{k},~~ \bm{\sigma}_1\cdot(\bm{q}\times\bm{k})\bm{\sigma}_2\cdot(\bm{q}\times\bm{k}).
\end{eqnarray}

In the static limit, Eq.~(\ref{Eq:covLO}) reduces to the LO chiral force in the HB scheme,
\begin{equation}
  V^\mathrm{HB} = (C_S+C_V) - (C_{AV}-2C_{T})\bm{\sigma}_1\cdot\bm{\sigma}_2  -\frac{g_A^2}{4f_\pi^2}\bm{\tau}_1\cdot\bm{\tau}_2
  \frac{\bm{\sigma}_1\cdot\bm{q}\bm{\sigma}_2\cdot\bm{q}}{\bm{q}^2 + m_\pi^2},
\end{equation}
which only contains the central, spin-spin and tensor interactions. It is important to note
that at LO the covariant power counting introduces three more LECs than the Weinberg approach and the modified Weinberg approach.

\subsection{Partial wave decomposition}

In this subsection, we follow the standard procedures given in Ref.~\cite{Erkelenz:1974uj} and evaluate the potentials
in the $LSJ$ basis, where $L$ denotes the total orbital angular momentum, $S$ the total spin, and $J$ the total angular
momentum. First, one calculates the matrix elements of the relativistic potential in the helicity basis, then one rotates them to the total angular momentum space $|JM\rangle$ with the help of Wigner $d$-functions. Finally, one transforms them to the $LSJ$ basis in terms of the Clebsch-Gordon coefficients. Below, we present the contact potential $V_\mathrm{CTP}$ contributing to different partial waves in the $|LSJ\rangle$ basis,

Interesting consequences can be seen in the contributions of $V_\mathrm{CTP}$ to different partial waves in the $|LSJ\rangle$ basis,
\begin{eqnarray}\label{Eq:VLSJfull}
  V_{1S0}&=& \xi_{N}\left[ C_{1S0} \left(1+R_p^2R_{p'}^2\right) + \hat{C}_{1S0}\left(R_p^2+R_{p'}^2\right)\right],\nonumber\\
  V_{3P0}&=& - 2\xi_N C_{3P0} R_pR_{p'},\nonumber\\
  V_{1P1}&=& - \frac{2\xi_{N}}{3} C_{1P1} R_pR_{p'}, \nonumber\\
   V_{3P1}&=& - \frac{4\xi_{N}}{3} C_{3P1} R_pR_{p'}, \nonumber\\
  V_{3S1}&=&\frac{\xi_{N}}{9}\left[ C_{3S1} \left(9+R_p^2R_{p'}^2\right)  + \hat{C}_{3S1} \left(R_p^2+R_{p'}^2\right)\right],\nonumber\\
  V_{3D1}&=& \frac{8\xi_{N}}{9}C_{3S1} R_p^2R_{p'}^2,\nonumber\\
  V_{3S1-3D1} &=& \frac{2\sqrt{2}\xi_{N}}{9}\left(C_{3S1} R_p^2R_{p'}^2 + \hat{C}_{3S1} R_p^2\right),\nonumber\\
  V_{3D1-3S1} &=& \frac{2\sqrt{2}\xi_{N}}{9}\left(C_{3S1} R_p^2R_{p'}^2 + \hat{C}_{3S1} R_{p'}^2\right),
\end{eqnarray}
where $\xi_{N}=4\pi N_p^2 N_{p'}^2$, $R_p=|\bm{p}|/\epsilon_p$, and $R_{p'}=|\bm{p}'|/\epsilon_{p'}$. The seven combinations of  $C_{S,A,V,AV,T}$ are
\begin{eqnarray}
C_{1S0} &=&(C_S+C_V+3C_{AV}-6C_T),\nonumber\\
\hat{C}_{1S0}&=&(3C_V +C_A+C_{AV}+6C_T),\nonumber\\
C_{3P0}&=&(C_S-4C_V+C_A-4C_{AV}),\nonumber\\
C_{1P1}&=&(C_S + C_A),\nonumber\\
C_{3P1}&=&(C_S-2C_V-C_A+2C_{AV}+4C_T),\nonumber\\
C_{3S1}&=&(C_S+C_V-C_{AV}+2C_T),\nonumber\\
\hat{C}_{3S1}&=&3 (C_V -C_A-C_{AV}+2C_T).
\end{eqnarray}
 $V_\mathrm{CTP}$ contributes to all partial waves with $J=0$, $1$, different from the (modified) Weinberg approach, where the contact terms only contribute to the $^1S_0$ and $^3S_1$ partial waves. The LO relativistic corrections in $V_{1S0}$ and $V_{3P0}$ have the same form as those introduced in the ``renormalization group invariant'' formulation~\cite{Nogga:2005hy,Birse:2005um,Valderrama:2009ei,Long:2012ve}.

For the OPEP, one can repeat the above procedure to obtain the partial wave  potentials for all angular momenta $J \geq
0$. Besides, in order to include  the retardation effect in the OPEP, consistent with the assumption of the Kadyshevsky equation, the following two types of integrals are needed,
containing the Legendre polynomials $P_J$,
\begin{equation}
  \int_{-1}^{+1} d z \frac{P_J(z)}{(E_{p'}-E_{p})^2-(\bm{p}'-\bm{p})^2-m_\pi^2} = -\frac{1}{|\bm{p}||\bm{p'}|} Q_J(z_\pi),
\end{equation}

\begin{equation}
  \int_{-1}^{+1} d z \frac{z P_J(z)}{(E_{p'}-E_{p})^2-(\bm{p}'-\bm{p})^2-m_\pi^2} = -\frac{1}{|\bm{p}||\bm{p'}|} Q^{(1)}_J(z_\pi),
\end{equation}
where $z$ denotes the cosine of the angle between $\bm{p}$ and $\bm{p}'$. $Q_J(z_\pi)$ is the Legendre function of the second kind, and $Q_J^{(1)}(z_\pi)= z_\pi Q_J(z_\pi) - \delta_{J0}$ with $z_\pi=(E_pE_{p'}-m_N^2+1/2m_\pi^2)/(|\bm{p}||\bm{p}'|)$.

With these integrals, the partial wave potentials of $V_\mathrm{OPEP}$ read as
\begin{itemize}
\item the spin singlet state:
\begin{equation}\label{Eq:OPEJ}
  V_{JJ}^{0J} = \frac{\pi g_A}{f_\pi^2} \left(Q_J^{(1)}(z_\pi) - \frac{E_pE_{p'}-m_N^2}{|\bm{p}||\bm{p}'| }Q_J(z_\pi)\right).
\end{equation}

\item the uncoupled spin triplet state:
\begin{equation}
  V_{JJ}^{1J} = \frac{\pi g_A}{f_\pi^2} \left(\frac{E_pE_{p'}-m_N^2}{|\bm{p}||\bm{p}'|} Q_J(z_\pi) - \frac{J+1}{2J+1}Q_{J-1}(z_\pi)-\frac{J}{2J+1}Q_{J+1}(z_\pi)\right).
\end{equation}
\item and the coupled triplet states:
\begin{equation}
  V_{J-1,J-1}^{1J} = \frac{\pi g_A}{(2J+1)f_\pi^2}\left[
  \frac{E_pE_{p'}-m_N^2}{|\bm{p}||\bm{p}'|}\left(-JQ_J^{(1)}(z_\pi) +\frac{(J+1)^2}{2J+1}Q_{J-1}(z_\pi)+\frac{J(J+1)}{2J+1}Q_{J+1}(z_\pi)\right) - Q_J(z_\pi)\right],
\end{equation}

\begin{equation}
  V_{J+1,J+1}^{1J} = \frac{\pi g_A}{(2J+1)f_\pi^2}\left[
  \frac{E_pE_{p'}-m_N^2}{|\bm{p}||\bm{p}'|}\left(-(J+1)Q_J^{(1)}(z_\pi) +\frac{J(J+1)}{2J+1}Q_{J-1}(z_\pi)+\frac{J^2}{2J+1}Q_{J+1}(z_\pi)\right) + Q_J(z_\pi)\right],
\end{equation}

\begin{eqnarray}
  V_{J-1,J+1}^{1J} &=& \frac{\pi g_A}{f_\pi^2}\frac{\sqrt{J(J+1)}}{2J+1}
  \left[
  \frac{E_pE_{p'}-m_N^2}{|\bm{p}||\bm{p}'|}\left(Q_J^{(1)}(z_\pi) +\frac{J+1}{2J+1}Q_{J-1}(z_\pi)+\frac{J}{2J+1}Q_{J+1}(z_\pi)\right)\right.\nonumber\\
  &&\left. - \frac{(E_p-E_{p'})m_N}{|\bm{p}| |\bm{p}'|} (Q_{J+1}(z_\pi)-Q_{J-1}(z_\pi)) - 2Q_J(z_\pi)
   \right],
\end{eqnarray}

\begin{eqnarray}\label{Eq:OPEJp1m1}
  V_{J+1,J-1}^{1J} &=& \frac{\pi g_A}{f_\pi^2}\frac{\sqrt{J(J+1)}}{2J+1}
  \left[
  \frac{E_pE_{p'}-m_N^2}{|\bm{p}||\bm{p}'|}\left(Q_J^{(1)}(z_\pi) +\frac{J+1}{2J+1}Q_{J-1}(z_\pi)+\frac{J}{2J+1}Q_{J+1}(z_\pi)\right)\right.\nonumber\\
  &&\left. - \frac{(E_p-E_{p'})m_N}{|\bm{p}| |\bm{p}'|} (Q_{J-1}(z_\pi)-Q_{J+1}(z_\pi)) - 2Q_J(z_\pi)
   \right].
\end{eqnarray}
\end{itemize}

In order to compare with the LO potential from the (modified) Weinberg power counting, we decompose the  relativistic potential into the sum of the static contribution and the relativistic corrections. For instance, the
$^1S_0$ partial wave potential [Eq.~(\ref{Eq:VLSJfull})],  expanded in terms of  $1/M_N$, reads
\begin{equation}\label{Eq:1S0VLSJ}
  V_{1S0}=4\pi \left[ C_{1S0} + (C_{1S0}+\hat{C}_{1S0})\left(\frac{\bm{p}^2+\bm{p}'^2}{4M_N^2} + \cdots \right)\right]+\frac{\pi g_A^2}{2f_\pi^2} \int_{-1}^{1}\frac{dz}{\bm{q}^2+m_\pi^2} \left[ \bm{q}^2 -
\left(
\frac{(\bm{p}^2-\bm{p}'^2)^2}{4M_N^2}+\cdots\right)\right].
\end{equation}
It is easy to single out the static contributions because the relativistic corrections are suppressed by $1/M_N^{2n}$
($n=1,2,\cdots$). In the covariant power counting, this argument is only true for the OPEP, where the same coefficient, $\pi
g_A^2/(2f_\pi^2)$, multiplies both the static contribution and the relativistic corrections. However, the situation for
the contact interaction is different: an independent LEC, $(C_{1S0}+\hat{C}_{1S0})$, determines the relativistic corrections of the CTP. Here, the $1/M_N$ expansion of Eq.~(\ref{Eq:1S0VLSJ}) is shown simply for the purpose of
comparison. In the numerical evaluation, we have used the relativistic interactions given in Eq.~(\ref{Eq:VLSJfull}) and Eqs.~(\ref{Eq:OPEJ}-\ref{Eq:OPEJp1m1}), where the retardation effect of the OPEP is taken into account.

\subsection{Scattering equation and phase shifts}
In order to calculate the partial wave $T$-matrix elements, the projected Kadyshevsky equation with  specific $LSJ$ can be written as
\begin{eqnarray}\label{eq:kadyshevsky}
   T^{SJ}_{L',L}(\bm{p}', \bm{p}) &=& V^{SJ}_{L',L}(\bm{p}', \bm{p}) \nonumber\\
   && + \sum\limits_{L''}
   \int_0^{+\infty}\frac{\bm{k}^2 d k}{(2\pi)^3} V^{SJ}_{L',L}(\bm{p}', \bm{k})
   \frac{M_N^2}{2(\bm{k}^2+M_N^2)} \frac{1}{\sqrt{\bm{p}^2+M_N^2}-\sqrt{\bm{k}^2+M_N^2} + i\epsilon} T_{L'',L}^{SJ}(\bm{k}, \bm{p}).
\end{eqnarray}
Furthermore, to remove ultraviolet divergences and to facilitate numerical calculations,
the potential has to be regularized.
Here, we choose the commonly used separable cutoff function~\cite{Epelbaum:1999dj},
\begin{equation}\label{Eq:formfactor}
  V_\mathrm{LO} \rightarrow V_\mathrm{LO}^\mathrm{Reg.} =
  V_\mathrm{LO}~ \mathrm{exp}\left(\frac{-\bm{p}^{2n}-\bm{p}'^{2n}}{\Lambda^{2n}}\right),
\end{equation}
with $n=2$. One should note that Eq.~(\ref{Eq:formfactor}) is not a covariant cutoff function. Although there are covariant cutoff functions of $q^2$,  they are not favored in constructing chiral forces because they will introduce additional angular dependence to partial
  wave potentials and thus affect the interpretation of contact interactions~\cite{Nogga:2005hy,Epelbaum:2000kv}. In the
  future, it would be interesting to construct a separable but covariant cutoff function and study the consequences. We
  also note that  in Ref.~\cite{Epelbaum:2014efa} an appropriate regularization method of the long-range interaction is
  applied to construct the chiral nuclear force. It would be interesting to apply such a prescription to our relativistic chiral force as well.


The partial wave $S$ matrix is related to the on-shell $T$ matrix by
\begin{equation}
  S_{L'L}^{SJ}(\bm{p}_\mathrm{cm}) = \delta_{L'L} + 2\pi i \rho T_{L'L}^{SJ}(\bm{p}_\mathrm{cm}),\quad \rho=- \frac{1}{16\pi^3} \frac{|\bm{p}_\mathrm{cm}| M_N^2}{E_p},
\end{equation}
where $\bm{p}_\mathrm{cm}$ is the C.M. three-momentum of the two-nucleon system. The phase space factor $\rho$ is determined by the elastic unitarity of the relativistic scattering equation, in this case the Kadyshevsky equation.
For the uncoupled cases, the phase shifts $\delta_{L}^{SJ}$ can be obtained from the on-shell $S$ matrix,
\begin{equation}
  S_{JJ}^{0J} = \exp(2i\delta_J^{0J}), \quad S_{JJ}^{1J} = \exp(2i\delta_J^{1J}),
\end{equation}

In order to calculate the phase shifts in the coupled channels ($J>0$), we use the ``Stapp''- or ``bar''- phase shift parametrisation~\cite{Stapp:1956mz} of the $S$ matrix, which can be written as
\begin{eqnarray}
  S &=& \left(
       \begin{array}{cc}
         S_{--}^{1J} & S_{-+}^{1J} \\
         S_{+-}^{1J} & S_{++}^{1J} \\
       \end{array}
     \right)\nonumber\\
   &=& \left(
       \begin{array}{cc}
         \exp(i\delta_-^{1J}) & 0 \\
         0 & \exp(i\delta_+^{1J}) \\
       \end{array}
     \right)
  \left(
       \begin{array}{cc}
         \cos(2\epsilon_J)  & i\sin(2\epsilon_J) \\
         i\sin(2\epsilon_J) & \cos(2\epsilon_J) \\
       \end{array}
     \right)
     \left(
       \begin{array}{cc}
         \exp(i\delta_-^{1J}) & 0 \\
         0 & \exp(i\delta_+^{1J}) \\
       \end{array}
     \right),
\end{eqnarray}
where the subscript ``$+$'' is $J+1$, ``$-$'' for $J-1$. The resulting phase shifts and mixing angles are
\begin{equation}
  \tan(2\delta_{\pm}^{1J}) = \frac{\mathrm{Im}(S_{\pm\pm}^{1J}/\cos(2\epsilon_J))} {\mathrm{Re}(S_{\pm\pm}^{1J}/\cos(2\epsilon_J))}, \quad
  \tan(2\epsilon_J) = \frac{-iS_{+-}^{1J}}{\sqrt{S_{++}^{1J}S_{--}^{1J}}}.
\end{equation}

\section{Results and discussion}

Numerically, we perform a simultaneous fit to the $J=0,~1$ Nijmegen partial wave phase shifts of the $np$ channel at laboratory kinetic energy~($E_\mathrm{lab}$)~\cite{Stoks:1993tb} values of $1$, $5$, $10$, $25$, $50$, and $100$ MeV. We do not take into account the errors of the phase shifts in the fit-$\tilde{\chi}^2$, defined as $\tilde{\chi}^2 = (\delta_\mathrm{LO}-\delta_\mathrm{PWA})^2$, mainly because the low energy partial wave phase shifts have very small uncertainties compared to the higher chiral order contributions neglected in our LO study.  In the present work, the pion and nucleon masses are fixed at $m_\pi=138.00$~MeV and $M_N=938.92$~MeV. The momentum cutoff $\Lambda$ is varied between $500$ MeV and $1000$ MeV.

\begin{figure}[t]
\includegraphics[width=0.48\textwidth,keepaspectratio,angle=0,clip]{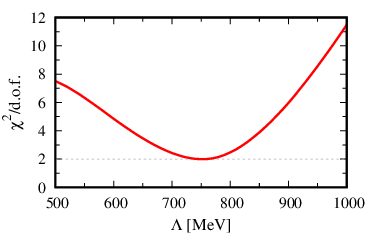}
\caption{Dependence of $\tilde{\chi}^2/\mathrm{d.o.f.}$ as a function of the momentum cutoff $\Lambda$.  The fit is to the $J=0,~1$ partial wave shifts of the $np$ channel with $E_\mathrm{lab}=1$, $5$, $10$, $25$, $50$, and $100$ MeV~\cite{Stoks:1993tb}.
\label{Fig:La-chi}
}
\end{figure}

The best fit result of $\tilde{\chi}^2/\mathrm{d.o.f.}$ is shown in Fig.~\ref{Fig:La-chi} as a function of the momentum
cutoff $\Lambda$. The minimum of $\tilde{\chi}^2/\mathrm{d.o.f.}$, $\sim2.0$, appears at
$\Lambda=750$ MeV.
The corresponding LECs $C_{S,A,V,AV,T}$  are listed in Table~\ref{Tab:LEC} and they are of similar
magnitude. The cutoff dependence indicates that the LO relativistic chiral force is not
renormalization group invariant. In the following discussion, although we take $\Lambda=750$ MeV as our
  relativistic result, the variance with the cutoff changing from 500 MeV to 1000 MeV should also be taken into
  account.

\begin{table*}[b]
\caption{Values of the leading order LECs (in unit of $10^4$GeV$^{-2}$) from the best fit (see text for details).
\label{Tab:LEC}}
\smallskip
\begin{ruledtabular}
\begin{tabular}{@{\extracolsep{\fill}}cccccc}
\noalign{\smallskip}
 LECs & $C_S$  & $C_A$ & $C_V$ & $C_{AV}$ & $C_T$
\smallskip
 \\
\hline\\[-0.05em]
Best fit & $-0.125$ & $0.040$ & $0.248$ & $0.221$ & $0.059$\\
\end{tabular}
\end{ruledtabular}
\end{table*}

\begin{figure}[t]
\includegraphics[width=0.49\textwidth,keepaspectratio,angle=0,clip]{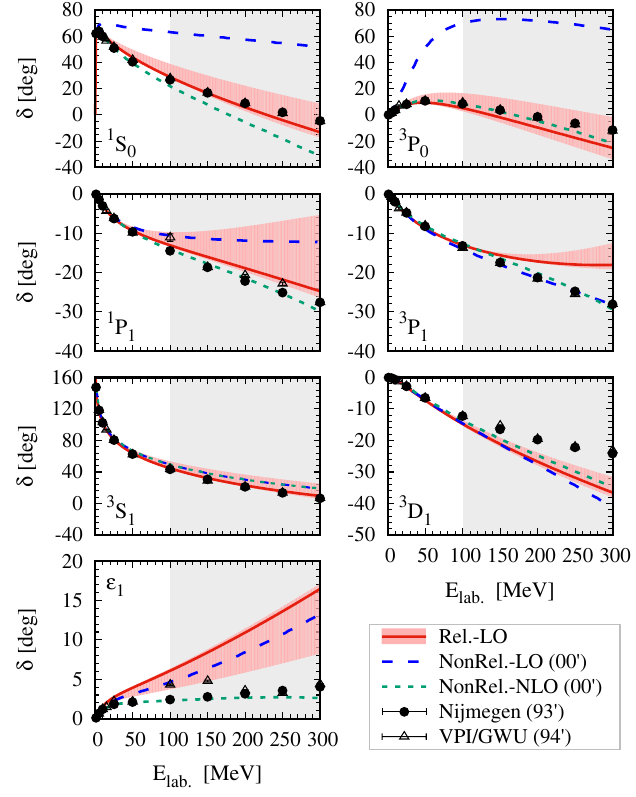}
\caption{Neutron-proton phase shifts for $J\leq1$. The red solid lines denote the best fitting results from the relativistic chiral NN potential, while the dashed and dotted lines represent the LO and NLO non-relativistic results respectively~\cite{Epelbaum:1999dj}. The red bands are the relativistic results with the cutoff ranging from 500 MeV to 1000 MeV. Solid dots and open triangles represent the $np$ phase shift analyses of Nijmegen~\cite{Stoks:1993tb} and VPI/GWU~\cite{Arndt:1994br} respectively. The gray backgrounds denote the energy regions where the  theoretical results are predictions.
\label{Fig:PSdes}
}
\end{figure}

With the best fit LECs, the description of the Nijmegen multi-energy~\cite{Stoks:1993tb} and the VPI/GWU
single-energy~\cite{Arndt:1994br} $np$ phase shifts up to $E_\mathrm{lab.}=300$ MeV are shown in Fig.~\ref{Fig:PSdes}.
The data of the latter analysis are not included in our fits. For comparison, the non-relativistic results~\cite{Epelbaum:1999dj} up to LO and NLO with $\Lambda=500$ MeV are also given in Fig.~\ref{Fig:PSdes}. Furthermore, the variations from the best fit results with the cutoff ranging from 500 MeV to 1000 MeV are shown as the red bands in the figure.
The relativistic formulation can improve the description of the phase shifts of $^1S_0$ and $^3P_0$ in
comparison with the LO non-relativistic results. The results of the LO relativistic chiral force are similar to those of the NLO non-relativistic chiral force. Furthermore, the variation of the cutoff does not qualitatively change  the overall picture.
The best description of the $^1P_1$ wave is slightly better than the NR counterpart, while the result for $^3P_1$ is slightly worse in the high energy region. For  the coupled $^3S_1$-$^3D_1$ waves, the LO relativistic and non-relativistic results are quantitatively similar when the cutoff variation is taken into account.

To understand the improvement in the two $J=0$ waves, we take the $^1 S_0$ channel as an example. The largest relativistic correction of the CTP is of the form
\begin{equation}
- \frac{\pi}{M_N^2}(C_{1S0}+\hat{C}_{1S0})\left(\bm{p}^2+\bm{p}'^2\right).
\end{equation}
This momentum-dependent term is desired to achieve a reasonable description of the $^1S_0$ channel for momenta around $m_\pi$, as shown in Refs.~\cite{Soto:2007pg,Long:2013cya}, where the $(\bm{p}^2+\bm{p}'^2)$ term is promoted on phenomenological grounds and the dibaryon field is introduced to deal with its resummation. This term has the same form as the NLO contribution of the NR chiral potential~\cite{Epelbaum:1999dj}. Lorentz invariance rearranges some of the higher order contributions in the NR potential to leading order in the relativistic potential. This mechanism is also behind the improved description of the $^3P_0$ partial wave phase shifts, where one contact term exists in the $V_{3P0}$ at LO, which has the same form as the NR NLO potential.

For the $^1P_1$ and $^3P_1$ partial waves, although, there is one similar contact term as the $^3P_0$ case, the description of phase shifts is almost the same as in the LO NR case. The reason is that the one-pion-exchange contribution could already describe  the $^1P_1$ and $^3P_1$ partial waves rather well. Numerically, we find that the contribution from the contact term is rather small.

For the coupled $^3S_1$-$^3D_1$ partial waves, the relativistic corrections are much more suppressed. For instance, the relativistic correction of $V_{3D1}$ is suppressed by at least $1/M_N^4$. As a result, the descriptions of these partial waves are similar to those of the LO Weinberg approach.

\begin{figure}[t]
\includegraphics[width=0.49\textwidth,keepaspectratio,angle=0,clip]{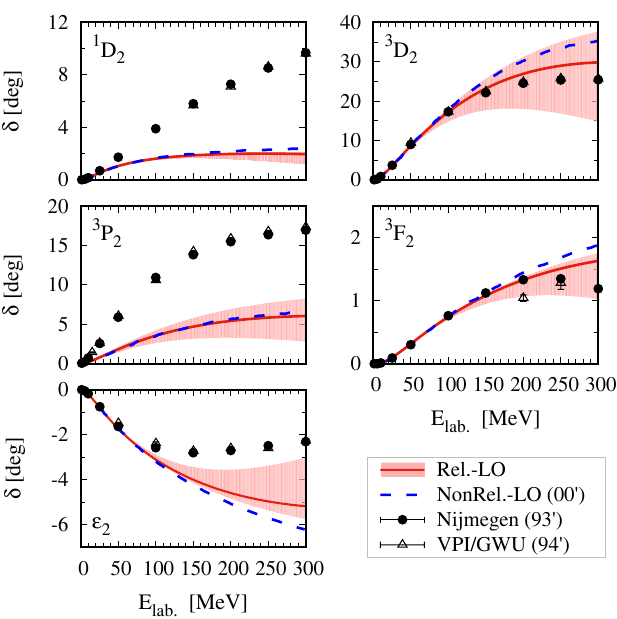}
\caption{Neutron-proton phase shifts for $J=2$. The notations are the same as Fig.~\ref{Fig:PSdes}.
\label{Fig:PSdesJ2}
}
\end{figure}
Furthermore, in Fig.~\ref{Fig:PSdesJ2}, we present the description of the $J=2$ phase shifts, where only one-pion-exchange diagrams contribute. Using the same notation as Fig.~\ref{Fig:PSdes}, we study the results obtained with $\Lambda=750$ MeV as a central value and the variation bands obtained with the cutoff varying from $500$ MeV to $1000$ MeV. The non-relativistic LO results from Ref.~\cite{Epelbaum:1999dj} are also shown for comparison. We can see that they are almost the same, as expected from Eq.~(\ref{Eq:1S0VLSJ}), where the relativistic corrections of the OPEP are largely suppressed.

Finally, using the LECs of Table~\ref{Tab:LEC}, we predict the binding energy of the deuteron to be $B_d=2.07$ MeV, which differs from its experimental value $B_d^\mathrm{exp}=2.22$ MeV by about 7\%. The scattering lengths of $^1S_0$ and $^3S_1$ turn out to be $a_{1S0}=-20.2$ fm and $a_{3S1}=5.6$ fm,  differing from their experimental counterparts, $-23.7$ fm and $5.4$ fm, by 15\% and 4\%, respectively.

\section{Summary and conclusion}
We have explored a new covariant power counting scheme to construct the nucleon-nucleon interaction in chiral effective field
theory. At leading order, the chiral force  includes part of the sub-leading terms  in the non-relativistic construction.
This force has been shown to lead to a  description of the Nijmegen partial wave phase shifts better than the LO Weinberg
approach and similar to the next-to-leading order Weinberg approach for the angular momenta $J=0$ partial waves.
For the higher waves, both approaches yield similar descriptions.
 Such an improvement of the description of phase shifts, even at leading order, encourages us to construct higher order relativistic chiral nuclear forces, which
may provide an essential input to relativistic nuclear structure studies.

In the present work, renormalization group invariance has not been achieved, as shown by the dependence of the results on the cutoff. In future, we would like to study this issue further, e.g., by modifying the
  power counting scheme. In addition, it will be interesting to study the convergence of the relativistic chiral force when higher order
  results become available. Furthermore, applications of the relativistic chiral force to nuclear matter using the Dirac Brueckner-Hartree-Fock theory
and to relativistic three-body problems using the approach proposed by H. Kamada et al.~\cite{Kamada:2002qt} are in progress.

 X.-L. R. acknowledges  valuable suggestions from and inspiring discussions with Evgeny Epelbaum.  We appreciate 
 discussions with  Ulf-G. Mei{\ss}ner, Manuel Pavon Valderrama, and Tetsuo Hyodo. This work was supported in part by the
 National Natural Science Foundation of China under Grants  No. 11375024, No. 11522539, No. 11335002,
 and No. 11375120, by DFG and NSFC through funds provided to the
Sino-German CRC 110 ``Symmetries and the Emergence of Structure in QCD'' (NSFC Grant
No. 11621131001, DFG Grant No. TRR110), the Major State 973 Program of China under Grant No. 2013CB834400, the China Postdoctoral Science Foundation under Grants No. 2016M600845, No. 2017T100008, the Fundamental Research Funds for the Central Universities, and by the DFG cluster of excellence \textquotedblleft Origin and Structure of the Universe\textquotedblright\ (www.universe-cluster.de).

\bibliography{refs}
%

\clearpage

\end{CJK*}

\end{document}